\documentclass{aa}  
\usepackage{graphicx}
\usepackage{txfonts}
\usepackage{hyperref}  
\hypersetup{colorlinks=true,linkcolor=[rgb]{1.,0.2,0.2},citecolor=[rgb]{0.1,0.4,1.},filecolor=[rgb]{0.7,0.2,0.2},urlcolor=[rgb]{0.7,0.2,0.2}}
\usepackage{color}
\definecolor{blue}{rgb}{0., 0., 1}
\newcommand {\T}{Table\,}
\newcommand {\Sec}{Sec.\,}
\newcommand {\Fig}{Fig.\,}

\begin{document}

   \title{Mind the Gap between A2061 and A2067: \\ Unveiling new diffuse large-scale radio emission}

   \author{G.V. Pignataro
          \inst{1,2} \fnmsep\thanks{e-mail: \href{mailto:giada.pignataro2@unibo.it}{giada.pignataro2@unibo.it}}
          \and
          A. Bonafede\inst{1,2}
          \and 
          G. Bernardi\inst{2,3,4}
          \and
          M. Balboni\inst{5,6}
          \and
          F. Vazza\inst{1,2,7}
          \and
          R. J. van Weeren\inst{8}
          \and
          F. Ubertosi\inst{1, 12}
          \and
          R. Cassano\inst{2}
          \and
          G. Brunetti\inst{2}
          \and
          A. Botteon\inst{2}
          \and
          T. Venturi\inst{2,4}
          \and
          H. Akamatsu\inst{9,10}
          \and
          A. Drabent\inst{11}
          \and
          M. Hoeft\inst{11}
        }

   \institute{Dipartimento di Fisica e Astronomia, Universit\`a degli Studi di Bologna, via P. Gobetti 93/2, 40129 Bologna, Italy
        \and
             INAF -- Istituto di Radioastronomia, via P. Gobetti 101, 40129 Bologna, Italy
        \and
            Department of Physics and Electronics, Rhodes University, PO Box 94, Makhanda, 6140, South Africa
        \and
            South African Radio Astronomy Observatory (SARAO), Black River Park, 2 Fir Street, Observatory, Cape Town, 7925, South Africa
        \and
            INAF – IASF Milano, Via A. Corti 12, 20133 Milano, Italy
        \and
            DiSAT, Università degli Studi dell’Insubria, Via Valleggio 11, 22100 Como, Italy
        \and
            Hamburger Sternwarte, Universitat Hamburg, Gojenbergsweg 112, D-21029, Hamburg, Germany 
        \and    
            Leiden Observatory, Leiden University, PO Box 9513, 2300 RA Leiden, The Netherlands
        \and 
            International Center for Quantum-field Measurement Systems for Studies of the Universe and Particles (QUP), The High Energy Accelerator Research Organization (KEK), 1-1 Oho, Tsukuba, Ibaraki 305-0801, Japan
        \and 
            SRON Netherlands Institute for Space Research, Utrecht, The Netherlands
        \and 
            Thuringer Landessternwarte, Sternwarte 5, D-07778 Tautenburg, Germany
        \and
            INAF—Osservatorio di Astrofisica e Scienza dello Spazio (OAS), via Gobetti 101, I-40129 Bologna, Italy}
   \date{Received September 15, 1996; accepted March 16, 1997}

 
  \abstract
   {}
   {The clusters Abell 2061 and Abell 2067 in the Corona Borealis supercluster have been studied at different radio frequencies and are both known to host diffuse radio emission. The aim of this work is to investigate the radio emission in-between them, suggested by low resolution observations.}
   {We analyse deep LOw Frequency ARray (LOFAR) High Band Antenna (HBA) observations at 144~MHz to follow-up on the possible inter-cluster filament suggested by previous 1.4~GHz observations. We investigate the radial profiles and the point-to-point surface brightness correlation of the emission in Abell 2061 with radio and X-ray observations, to describe the nature of the diffuse emission. }
   {We report the detection of diffuse radio emission on 800~kpc scale, more extended than previously known, reaching beyond the radio halo in Abell 2061 towards Abell 2067 and over the separation outside the two clusters $R_{500}$ radii. We confirm the presence of a radio halo in A2061, while do not find evidence of diffuse emission in Abell 2067. The surface brightness profile from the centre of A2061 shows an excess of emission with respect to the azimuthally averaged radio halo profile and X-ray background. We explore three different dynamical scenario to explain the nature of the diffuse emission. Additionally, we analyse a trail of emission of $\sim760$~kpc between the radio halo and radio relic in Abell 2061.}
   {This dynamically-interacting, pre-merger system closely resembles the two other cluster pairs where radio bridges connecting the radio halos on Mpc-scales have been detected. The diffuse emission extends beyond each cluster $R_{500}$ radius but in this unique case, the absence of the radio halo in Abell 2067 is likely the reason for the observed `gap' between the two systems. However, the point-to-point correlation results are challenging to explain. The classification of the emission remains unclear, and detailed spectral analysis and further X-ray observations are required to understand the origin of the diffuse emission.}

  \maketitle
%

\section{Introduction}\label{sec:intro}

\begin{figure*}[h!]
    \centering
    \includegraphics[width=1\linewidth]{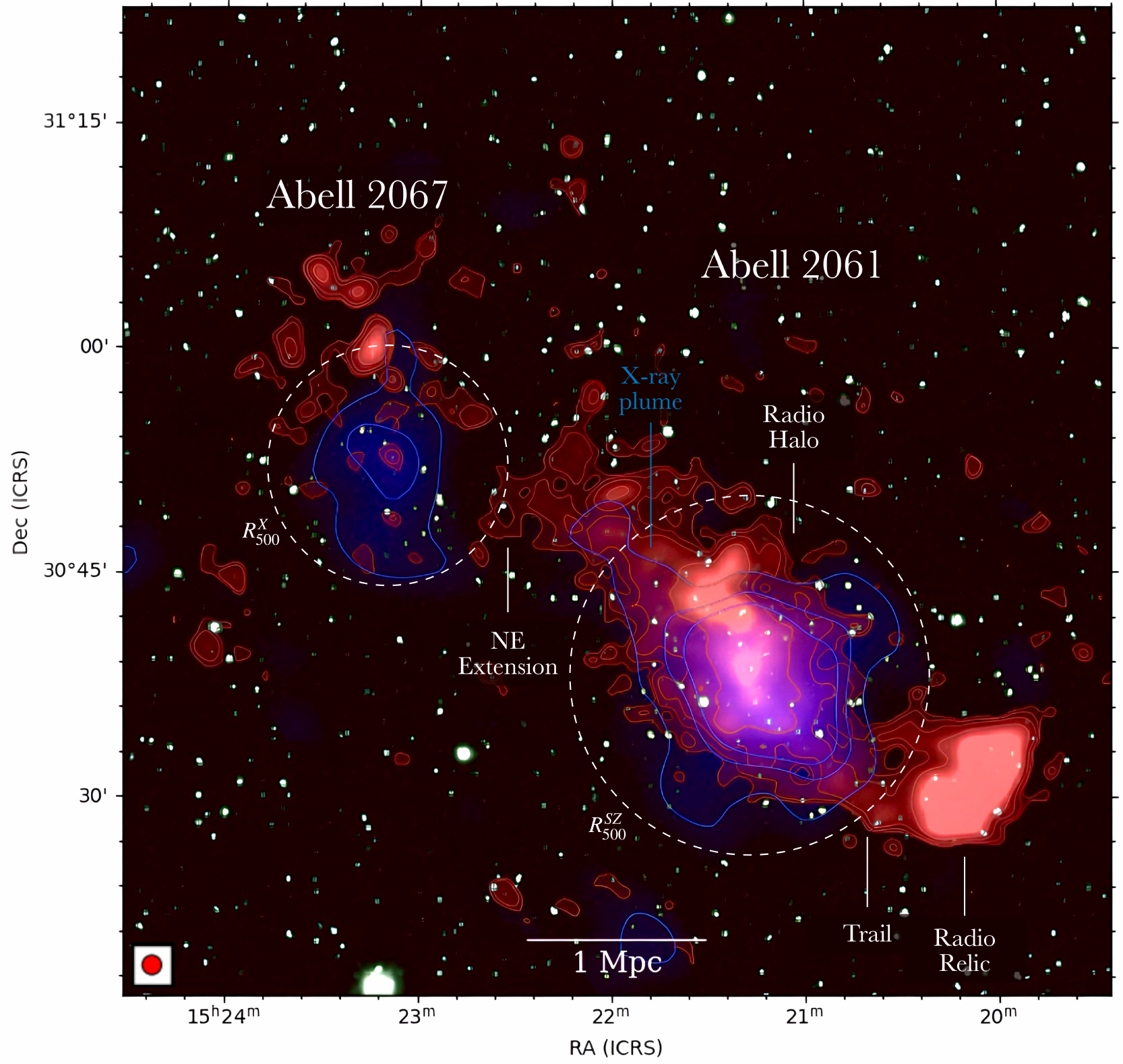}

    \caption{Composite image of the galaxy clusters system Abell 2061 - Abell 2067 in the Corona Borealis Supercluster. The radio emission from  LOFAR observations at 144~MHz and 80$''$ resolution is shown in red. The diffuse emission sources presented in this work are highlighted in white. The X-ray, ROSAT emission is shown in blue, overlaid on the optical, Pan-STARRS1 image. }
    \label{fig:composite}
\end{figure*}

Diffuse radio synchrotron emission is observed in merging galaxy clusters, tracing the presence of magnetic field and relativistic particles on scales of hundred to thousands of kpc. Cluster merger events are among the most energetic in the Universe, and a fraction of this energy is channeled into the re-acceleration of particles that produces diffuse radio emission of various morphologies. 
Radio halos are typically observed to be unpolarised and at the center of galaxy cluster, extending from hundreds of kpc to Mpc scales \citep{vanweeren2019}. Their morphologies can range from round to elongated, and present smooth or clumpy emission  \citep{bonafede2012, venturi2013}. The radio halo luminosity is usually well correlated with X-ray luminosity and temperature of the host cluster \citep{cassano2010,cassano2023}.
Unlike halos, radio relics are found at the peripheries of merging clusters, usually extending up to Mpc scales with an elongated morphology. Relics are often observed to be highly polarised, and trace the axis of the accretion process, where the resulting shocks reaccelerate particles and amplify the magnetic fields. 
Recently, low-frequency radio observations revealed the presence of diffuse emission on ever larger scales, tracing the regions connecting interacting galaxy clusters \citep{govoni2019, botteon2018, botteon2020, hoeft2021, pignataro2024a}. Other types of extended filamentary emission is detected between radio halos and relics \citep{pasini2022,bonafede2021} and between clusters and groups \citep{bonafede2021, venturi2022}. These observations suggest that magnetic fields and relativistic particles are a component of the large scale structure, and that the process of major merger can produce a favourable environmnent, where turbulence and shocks trigger particle re-acceleration on very large scale. 

Therefore, the conditions to generate diffuse radio emission on very large scales are expected to be particularly favourable in superclusters of galaxies, the largest coherent structures in the Universe, where rich clusters in their core may be dynamically active \citep{einasto2021}. 
The Corona Borealis supercluster (CSC) is the most prominent and dense supercluster in the Northern Sky, at an average redshift of $z\sim0.07$. It is composed of ten galaxy clusters, including which Abell 2056, Abell 2061, Abell 2065, Abell 2067, and Abell 2089 comprise a gravitationally-bound supercluster core that is collapsing \citep{pearson2014}. The dominant cluster of the CSC is Abell 2065 \citep{markevitch1999}, and recent studies have found Abell 2061-Abell 2067 (hereafter, A2061 and A2067) to be a gravitationally-bound pair, in a pre-merger state \citep{marini2004, batiste2013, pearson2014}. This system closely resembles the cluster pairs A399-A401 and A1758N-S, where radio bridges have been observed at low frequency \citep{govoni2019, botteon2020, dejong2022, pignataro2024a}. 

The two cluster centres are separated by a projected distance of $\sim$2.5~Mpc \citep{rines2006}, and A2061 is the main cluster of this pair. It has a $R^{SZ}_{500}\sim$1~Mpc radius, \citep{planck2016}
and an X-ray luminosity of $L_{X}^{0.1-2.4 keV}\sim 3.95 \times 10^{44}$ erg s$^{-1}$ \citep{ebeling1998}. Its X-ray disturbed morphology suggests it is undergoing mergers with the surrounding, smaller halos \citep{marini2004}. In particular, it contains an X-ray extension (dubbed \textit{plume} in \citealt{marini2004}) visible with ROSAT-Position Sensitive Proportional Counters (PSPC) extending towards A2067 in the North-East (NE) direction. However, since the interaction between the two clusters appears to be in early stages \citep{marini2004}, the plume is likely related to a third substructure that is currently interacting or has already interacted in the past with A2061. \cite{marini2004} suggested the interaction of the two clusters could be an indication of the existence of an underlying filament, along which also the group merged into A2061. 

A2067 is, in turn, a relatively low X-ray luminosity cluster ($L_{X}^{0.1-2.4 keV}\sim4\times 10^{43}$ erg s$^{-1}$), which also shows an elongated morphology in the ROSAT All Sky-Survey (RASS) image. The X-ray peak coordinates differ by $\sim8'$ from the ACO center, and the measured X-ray radius $R_{500}^{X}$ is $\sim0.7$ Mpc \citep{piffaretti2011}. The quantities related to the two clusters are summarized in \T\ref{tab:coord}.

\begin{table}[]
\renewcommand{\arraystretch}{1.3}
\centering
\resizebox{\columnwidth}{!}{%
\begin{tabular}{cccccc}
               & \textbf{\begin{tabular}[c]{@{}c@{}}RA \\ {[J2000]}\end{tabular}} & \textbf{\begin{tabular}[c]{@{}c@{}}Dec \\ {[J2000]}\end{tabular}} & \textbf{z} & \textbf{\begin{tabular}[c]{@{}c@{}}$R_{500}$ \\ {[}Mpc{]}\end{tabular}} & \textbf{\begin{tabular}[c]{@{}c@{}}$M_{500}$ \\ {[}$10^{14} M_{\odot}${]}\end{tabular}} \\ \hline \hline
        
\textbf{A2061} & 15h 21m 08s & +30°38'08''    & 0.0783     & 1.0                                                                     & 3.5                                                                                     \\
\textbf{A2067} & 15h 23m 07s & +30°50'42''    & 0.0756     & 0.7                                                                     & 1.2   \\ \hline                                                                                 
\end{tabular}%
}
\smallskip
\caption{Coordinates, redshift, radius and mass of the two clusters analysed in this work. The reported coordinates are referring to the X-ray centers, from XMM observations for A2061 \citep{bartalucci2023}, and ROSAT for A2067 \citep{piffaretti2011}. The quantities $R_{500}$ and $M_{500}$ are measured from SZ  \citep{planck2016} and X-ray \citep{piffaretti2011} observations for A2061 and A2067, respectively.}
\label{tab:coord}
\end{table}

Radio observations of this system have been carried out at different frequencies to investigate evidence of diffuse emission. For A2061,  \cite{kempner2001} and \cite{rudnick2009} reported a possible relic, detected at 1.4~GHz with the NRAO VLA Sky Survey \citep[NVSS,][]{condon1998} and 327~MHz with the Westerbork Northern Sky Survey \citep[WENSS,][]{rengelink1997}. The presence of the relic is then confirmed with WSRT observations at 1.38 and 1.7~GHz, which results in a $\alpha=-1.03\pm0.09$ spectral index \citep{vanweeren2011}. Additionally, \cite{rudnick2009} reported a tentatively detected radio halo at 327~MHz at the center of the cluster. The radio halo is also tentatively detected by GBT-NVSS observations at 1.4~GHz reported in \cite{farnsworth2013}, where they derive a preliminary spectral index for the halo of $\alpha=-1.8\pm0.3$ between 0.3 and 1.4~GHz. In the GBT-NVSS and WENSS images, the radio halo appears elongated, displaying an extension towards the NE likely associated with the X-ray plume. \cite{farnsworth2013} also reported the presence of a possible bridge of emission, joining the radio halo and the relic, and a possible inter-cluster filament between A2061 and A2067, seen at a limited statistical significance in images with $11'$ angular resolution at 1.4~GHz. However, the very poor resolution of GBT does not allow a firm conclusion on this classification. 
A2067 is also observed with GBT-NVSS at 1.4~GHz, where they detect a marginally resolved emission $\sim12'$ north of the X-ray peak. However, there appears to be a blending of unresolved emission and they classify this emission as a possible relic \citep{farnsworth2013}. Recently, A2061 was observed with the LOFAR Two-Metre Sky Survey \citep[LoTSS,][]{shimwell2017, shimwell2019, shimwell2022} and was studied as part of the Planck \citep{planck2016} clusters sample covered by the LoTSS-DR2 \citep{botteon2022}. They were able to confirm the presence of the radio halo, the radio relic, and the trail between them at 144~MHz. 

In this work, we present 16 hours new LOFAR High Band Antenna (HBA) observations at 144~MHz of the A2061-A2067 system, following up the LoTSS observations that already suggested the presence of a bridge of emission to the region of the X-ray plume. Here, we report the discovery of further extended diffuse emission between the two clusters (see \Fig\ref{fig:composite}) that might be related to the filament connecting the two clusters \citep{einasto2021}. 
This paper is organized as follows: in \Sec\ref{observations} we describe the radio and X-ray data reduction; in \Sec\ref{sec:radialss}
we compare radio and X-ray radial profiles in the direction of the extended radio feature; in \Sec\ref{sec:pointopoint} we investigate the radio/X-ray correlation for A2061; finally, in \Sec\ref{sec:discussion} we discuss the possibile dynamical scenarios to originate the diffuse emission and its classification. Throughout this work we assume a $\Lambda$CDM cosmology, with $H_{0}=70$ km s$^{-1}$~Mpc$^{-1}$, $\Omega_{m}=0.3$, and $\Omega_{\Lambda}=0.7$. With these assumptions, at the average distance of the A2061-A2067 system (z$\sim0.076$), $1^{\prime}=85$~kpc and the luminosity distance is $D_{\rm L}= 345$~Mpc.

\begin{figure*}[h!]
    \centering
    \includegraphics[width=1\linewidth]{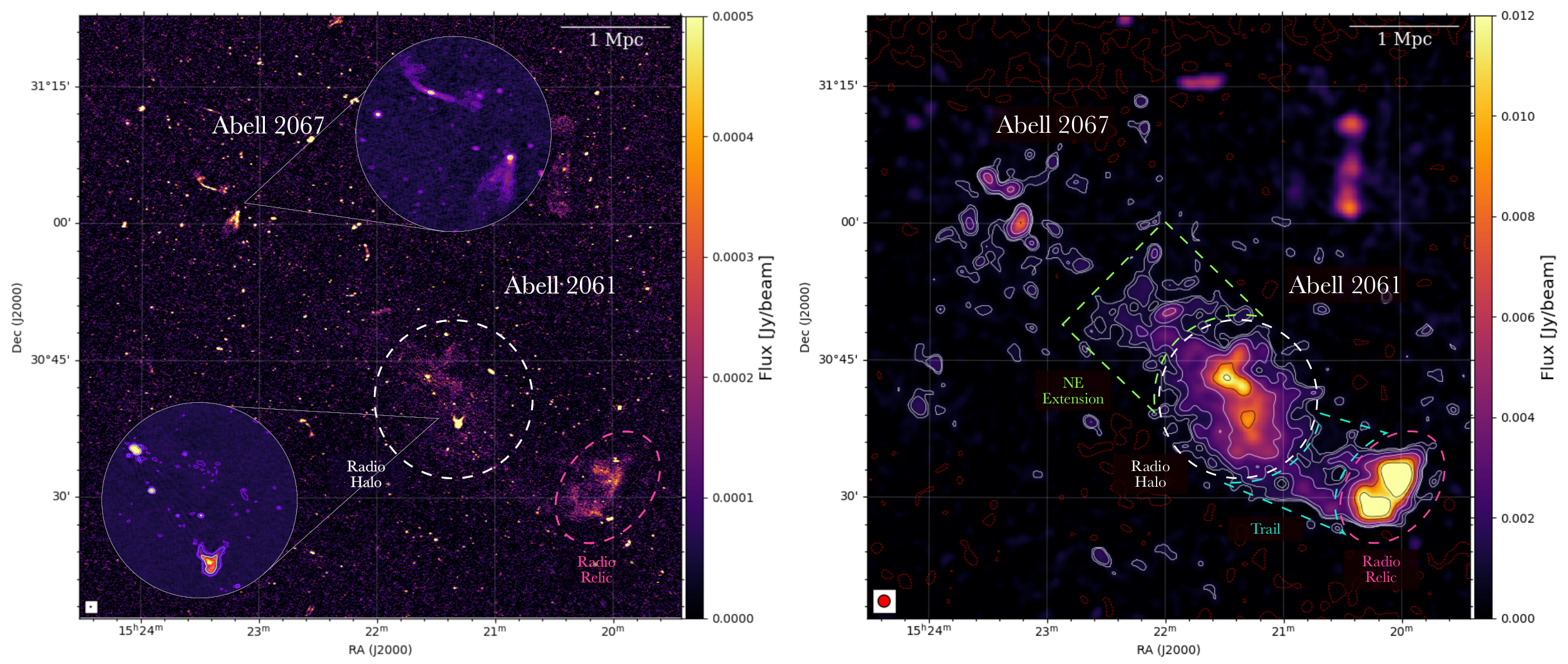}
    \caption{Radio maps of A2061-A2067. \textit{Left panel}: Image of A2061-A2067 at a resolution of $8''\times5''$ (p.a. $90$°). The imaging was done adopting a briggs weighting scheme \citep{Briggs} with \texttt{robust=}$-0.5$, and the resulting rms noise is $\sigma_{ rms}=80$ $\mu$Jy~beam$^{-1}$. The radio halo (white dashed circle) and the radio relic (magenta dashed circle) are already visibile at high resolution and without compact source subtraction. The most interesting radio sources in the field are highlighted in the circles. \textit{Right panel}: Image of A2061-A2067 at a resolution of $80''$, after subtraction of compact sources. The imaging was made adopting a briggs weighting scheme with \texttt{robust=}$-0.5$ and applying a Gaussian \textit{uv}-taper of $70''$. We show the $2,3,5,10-\sigma_{rms}$ contour levels in white, and the $20,30-\sigma_{rms}$ contours level in black ($\sigma_{rms}=0.4$ mJy~beam$^{-1}$.) The $-2\sigma_{rms}$ contours are shown in dashed red. Additionally to the radio halo and relic in A2061, we now detect diffuse emission in the NE extension (green dashed region) and trail (cyan dashed region).
    }
    \label{images}
\end{figure*}

\section{Observations and data reduction}\label{observations}
\subsection{Radio data}
The two clusters were observed as part of a pointing on the northern part of CSC, therefore the system is distant $\sim$35’ from the pointing center. We observed the CSC with LOFAR HBA using the same setup as LoTSS, that is in the frequency range 120-168 MHz, divided into 244 sub-bands of 64 channels each. The total observing time for CSC North was 32 hours (proposal code: LC014\_18, P.I: Vazza), however only 16 hours are not affected by severe ionospheric conditions and used for this analysis.
This observation was carried out in co-observing mode with the LoTSS and the data reduction follows the LoTSS reduction scheme. We summarize below the main steps and refer the reader to \cite{shimwell2022} and \cite{tasse2021} for further details. 
The pipeline performs direction-independent and dependent calibration and imaging of full field of view on CSC North, using \texttt{PREFACTOR} \citep{vanweeren2016, mechev2018, degasperin2019}, \texttt{KILLMS} \citep{tasse2014,smirnov2015} and \texttt{DDFACET}  \citep{tasse2018}.
To allow flexible re-imaging limited to the target of interest, we then performed the so-called 'extraction' procedure \citep{vanweeren2021}, where the direction-dependent solutions are used to subtract all sources outside a box of approximately 0.8° radius around the target.
As a final step, additional cycles of phase self-calibration in the extracted field are performed, followed by several rounds of amplitude calibration using a longer solution interval. After this last refinement, we used the final calibrated, extracted visibilities of the target to image at different resolutions with \texttt{WSClean v3.1} \citep{offringa2014}.\\ 
Throughout the paper, we report the measured flux densities $S_{\nu}$ of radio sources with uncertainties estimated as 
\begin{equation}
    \sigma_{S}= \sqrt{(S_\nu\cdot f)^{2}+N_{b}\cdot (\sigma_{\rm rms})^{2}},
\end{equation}
where $f=0.1$ is the absolute flux-scale uncertainty \citep{hardcastle2021, shimwell2022}, $N_{b}$ the number of beams covering the source, and $\sigma_{\rm rms}$ the rms noise sensitivity of the map.
The radio power $P_\nu$ is calculated from the flux density $S_\nu$ as \citep{condon1988}
\begin{equation}
    P_\nu = 4 \pi D_L^2 S_\nu (1+z)^{\alpha - 1}
\end{equation}
where $D_L$ is the luminosity distance, and the k-correction adjust for the redshift $z$ of the source with spectral index $\alpha$, defined by $S_\nu\propto\nu^{-\alpha}$.
\par
We produced a final primary beam corrected image at the central frequency of 144 MHz, at a resolution of $8''\times5''$ (p.a. $90$°) with a rms noise of $\sigma_{\rm rms}=80$ $\mu$Jy~beam$^{-1}$, shown in \Fig\ref{images} (left panel). From the high-resolution image, it is already possible to see the diffuse emission from the radio halo and relic in A2061, and a hint of emission connecting them.
Since we are particularly interested in the diffuse emission in the field, we produced a high-resolution image excluding baselines shorter than $\sim760\lambda$ (i.e. emission on scales more extended than $\sim4.5'\simeq380$ kpc) to recover only the compact sources, and then subtracted their components from the visibilities. Finally, a $80''$ resolution, source-subtracted image of the target field with a rms noise of $\sigma_{\rm rms}=0.4$ mJy~beam$^{-1}$ is shown in \Fig\ref{images} (right panel). In the A2067 field, no diffuse emission is revealed other than residuals from extended sources, such as the tailed Active-Galactic-Nuclei (AGN) visible in \Fig\ref{images} (left panel). In the A2061 field, the low-resolution image shows the presence of the radio halo and relic in A2061, detected at high significance. Additionally, we detect the trail of emission connecting the radio halo and relic, and reveal filamentary diffuse emission extending over $\sim800$ kpc from the radio halo in A2061 towards A2067, which we label as 'NE extension' in \Fig\ref{fig:composite} and \Fig\ref{images}. 

\subsection{X-ray data}

The galaxy cluster A2061 has been observed with XMM-Newton (Obs. id: 0721740101) and it also belongs to the Cluster HEritage project with \textit{XMM-Newton} - Mass Assembly and Thermodynamics at the Endpoint of structure formation \citep[CHEX-MATE,][]{chexmatecolab2021} project sample. For the aim of this study, it is useful to compare the properties of the thermal and non-thermal emission in A2061.

The X-ray data on A2061 were processed using the CHEX-MATE pipeline as detailed by \cite{bartalucci2023} and \cite{rossetti2024}. Below, we summarize the main steps involved.

Observations were conducted with the European Photon Imaging Camera \citep[EPIC, a set of three X-ray CCD cameras,][]{turner2001, struder2001} and the datasets were then reprocessed using the Extended-Science Analysis System \cite[E-SAS,][]{snowden2008}. Flare events were filtered out by excluding time intervals with count rates exceeding $3\sigma$ above the mean count rate. Point sources were also excluded from the analysis based on the methodology presented in \cite{ghirardini2019} and \cite{bartalucci2023}.
The scientific images of the cluster are generated from each camera's photon-count images in the [0.7-1.2]~keV band, which optimizes the signal-to-noise ratio for cluster thermal emission \citep{ettori2010}. Exposure and background maps are also created. The X-ray background consists of a sky component (i.e. the local Galactic emission and the Cosmic X-ray background) and an instrumental component introduced by the interaction between high-energy particles and the detector. Following \cite{ghirardini2019} and \cite{bartalucci2023}, the instrumental background component was then mitigated and the sky component described by a constant profile component.
Finally, to maximize statistics, the images, exposure maps, and background maps from the three cameras were merged \citep{bartalucci2023} and used for the following analysis. 

\section{Results and discussion}

In the next sections, we present and discuss the analysis on the radio and X-ray data performed primarily on the NE extension of diffuse emission discovered in A2061. We also discuss the several diffuse sources found in this cluster, whose discovery has been already reported in dedicated papers (see \Sec\ref{sec:intro}). We present the properties of these sources and add complementary information that we obtained with our new data. 

\subsection{Radial profiles}\label{sec:radialss}
\begin{figure*}[h!]
    \centering
    \includegraphics[width=1\linewidth]{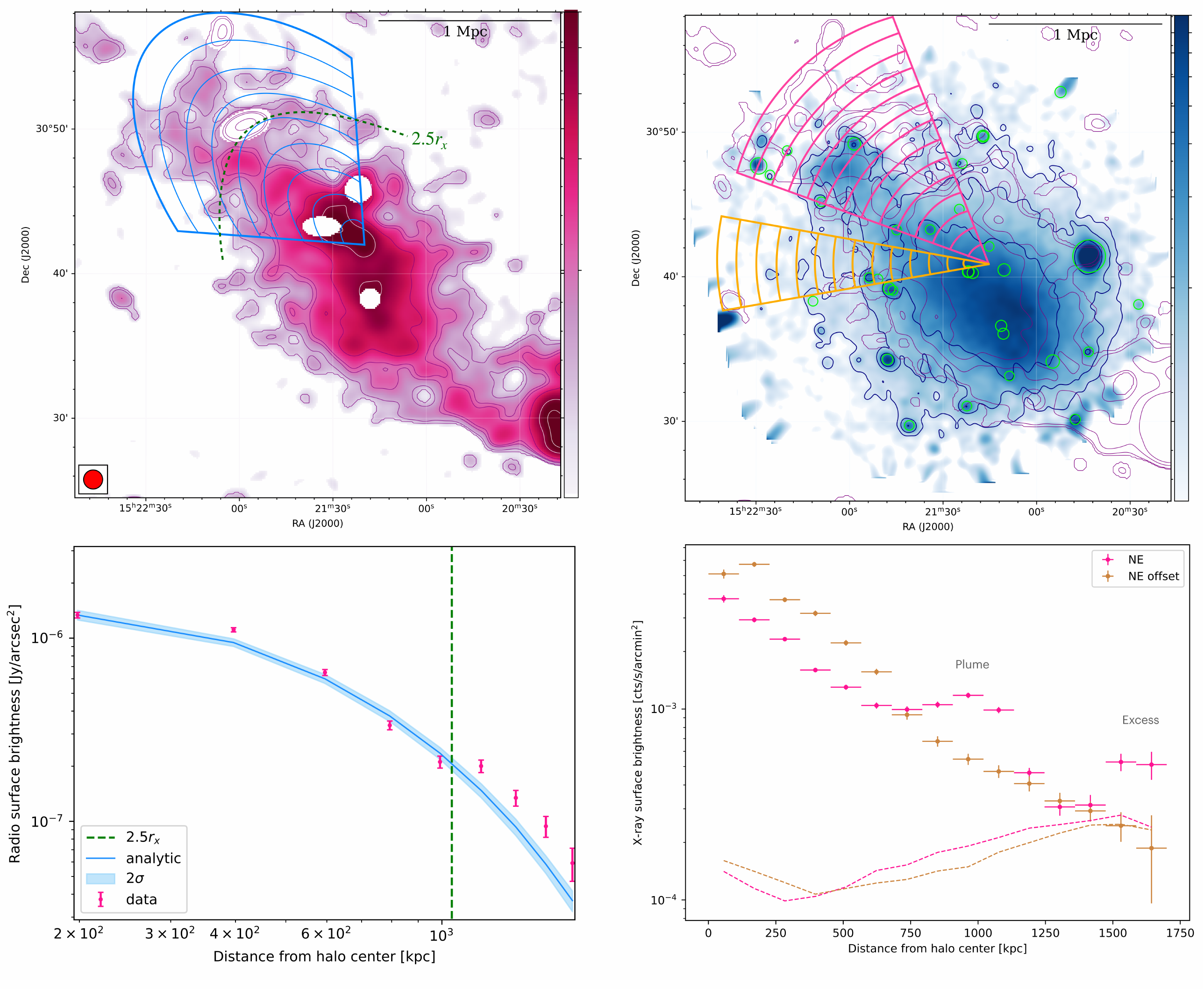}
    \caption{Radio and X-ray surface brightness profile of A2061. \textit{Top left panel}: the $80''$ radio map at 144~MHz with residual sources masked is shown in pink colourscale. We show the $2,3,5,10-\sigma_{rms}$ contour levels in purple, and the $20,30-\sigma_{rms}$ contours level in white. The mean radio brightness is computed inside the blue region. The width of the elliptical annuli along the major axis is $\sim 2$ times the beam size ($80''$). The $2.5r_{x}$ area is delimited by the dashed green line. \textit{Bottom left panel}: comparison between the average surface brightness measured in each blue bin (the magenta data points, with 1$\sigma$ uncertainties) with the model profile from the best-fit analysis performed with \texttt{HALO-FDCA} (solid cyan line, with 2$\sigma$ uncertainties). Beyond $\sim 2.5r_{x}$ (dashed green line) the average surface brightness is persistently higher than the best fit radio halo model, due to the NE extension. \textit{Top right panel}: the smoothed \textit{XMM-Newton} X-ray map is shown in blue colourscale. The contour-levels from the radio emission are overlaid in purple and the X-ray point sources (green circles) are masked. The X-ray surface brightness is computed inside the magenta slices, centered on the radio halo and covering the NE extension. We compare this sector with the yellow slice, in an offset direction from the extension. The width of the circular sectors is one beam size ($80''$). \textit{Bottom right panel}: The two profiles extracted from the different sectors are shown in magenta and yellow points. In the NE direction, we note a bump in the profile associated with the X-ray plume, that is followed by an excess at larger radii. On the contrary, there is no evidence of excess at the same distances from the center in the yellow sector. The background level in each sector is shown with magenta and yellow dashed lines.
    }
    \label{radial_profile}
\end{figure*}
\subsubsection{Radio}\label{sec:radioradial}

The most interesting source of diffuse emission in A2061 is the filamentary emission that is connected to the radio halo and extends in the direction of A2067, creating a bridge of emission between the two $R_{500}$ (\Fig\ref{fig:composite}).
In order to investigate the nature of the emission from the radio halo and the extended feature, we computed a radial profile using the radio map from our 144~MHz observations. The surface brightness profile of the radio halo in A2061 was already investigated in \cite{botteon2022} as part of the Planck (PSZ2 catalog, \citealt{planck2016}) clusters sample covered by the LoTSS-DR2. The radio profile was studied with the use of the Halo-Flux Density CAlculator (\texttt{HALO-FDCA\footnote{\hyperlink{https://github.com/JortBox/Halo-FDCA}{ https://github.com/JortBox/Halo-FDCA}}}; \citealt{boxelaar2021}). This code fits the two-dimensional brightness profile using Bayesian inference. The Markov Chain Monte Carlo (MCMC) algorithm is perfomed to determine the best-fit parameters and associated uncertainties. We repeat the analysis on A2061 on our $80''$ radio map, with additional masking of the residuals from poorly subtracted bright diffuse sources which differs from the masking performed in \cite{botteon2022}. The extension emission is fitted as part of the radio halo.
Typically, radio halos brightness profiles are fitted by an exponential law of the form:
\begin{equation}
    I_{\mathbf{r}}= I_{0} \, e^{-G(\mathbf{r})}
\end{equation}
where $I_{0}$ is the central radio surface brightness \citep{murgia2009}, $G(\mathbf{r})$ is a function that describes the model morphology (i.e. circular, elliptical, or skewed) and $\mathbf{r}$ is the positional vector.
The radio halo in A2061 was fitted with a rotated ellipse morphology that allows for a rotation $\phi$ with respect to the coordinate system centered in $(x_{0}, y_{0})$ \citep{boxelaar2021}. Therefore $G(r)$ becomes:
\begin{equation}
    G(\mathbf{r})= \left[\left(\frac{X_{\phi}(\mathbf{r})}{r_{x}}\right)^{2} + \left(\frac{Y_{\phi}(\mathbf{r})}{r_{y}}\right)^{2}\right]^{0.5}
\end{equation}
where $X_{\phi}$, $Y_{\phi}$ represent the coordinate transformation, and $r_{x}$ and $r_{y}$ represent the two \textit{e}-folding radii in the direction of the major and minor axis. This model has six free parameters: $I_{0}$, $x_{0}$, $y_{0}$, $r_{x}$, $r_{y}$, and $\phi$.
The best-fit parameters and their uncertainties are reported in \T\ref{tab:halo-fdca}. The residual from the fitting procedure are shown in Appendix \ref{appendix}. In \Fig\ref{radial_profile}, we show the radial profile of the halo brightness
at 144 MHz. We have computed the mean of the radio surface brightness and its error within an elliptical annuli having a width $\sim2.3'$ that covers the NE extension. The masked pixels are excluded when calculating the surface brightness. We compare the measured data with the analytical profile evaluated for the same annuli, computed with the model parameters. The best-fit parameters are Monte Carlo resampled $500$ times inside their uncertainties, and the final analytical mean radio brightness and its standard deviation are shown in \Fig\ref{radial_profile} (left panels). The data and the model profile show overall good agreement, with small deviations at small radii. The discrepancy between real and model data when fitting the radio halo with a single exponential profile were already noted in recent works \citep[see e.g.][]{cuciti2021, botteon2022}. For disturbed merging clusters as is the case of A2061, \cite{botteon2023} showed that, especially in the central regions, these deviations are likely related to the presence of substructure in the brightness distribution of the real halo. We note that after $\sim2.5r_{x}$, i.e. over the NE extension, the data points show a systematic excess of emission with respect to the radio halo profile. This result questions on the origin of the extended emission. The measured excess might indicate that the emission we observe in the NE extension has a different origin than the radio halo, and could be classified as a radio bridge. We will investigate these possibilities using X-ray data.

\begin{table}[]
\renewcommand{\arraystretch}{1.3}
\centering
\resizebox{\columnwidth}{!}{%
\begin{tabular}{ccllll}
\textbf{$\mathbf{I_{0}}$ {[}$\mu$Jy arcsec$^{2}${]}} & \textbf{$\mathbf{x_{0}}$ {[}deg{]}} & \textbf{$\mathbf{y_{0}}$ {[}deg{]}} & \textbf{$\mathbf{r_{x}}$ {[}kpc{]}} & \textbf{$\mathbf{r_{y}}$ {[}kpc{]}} & $\mathbf{\phi}$ \textbf{{[}rad{]}} \\ \hline
\textbf{$1.82$}                             & $230.33$                   & $30.68$           & $418.7$           & $212.7$           & $2.14$           \\
\textbf{$0.03$}                             & $0.0015$                   & $0.0017$          & $8.2$            & $4.4$             & $0.018$           \\ \hline
\end{tabular}%
}
\smallskip
\caption{\texttt{Halo-FDCA} best fit parameters (top row) and uncertainties (bottom row) for the radio halo in A2061. The model assumed to fit the radio halo is a rotated ellipse, which has six free parameters. The reduced $\chi^{2}$ of the best fit is $1.1$. }
\label{tab:halo-fdca}
\end{table}
\subsubsection{X-ray}\label{sec:xrayradial}
\begin{figure}[h!]
    \centering
    \includegraphics[width=1\linewidth]{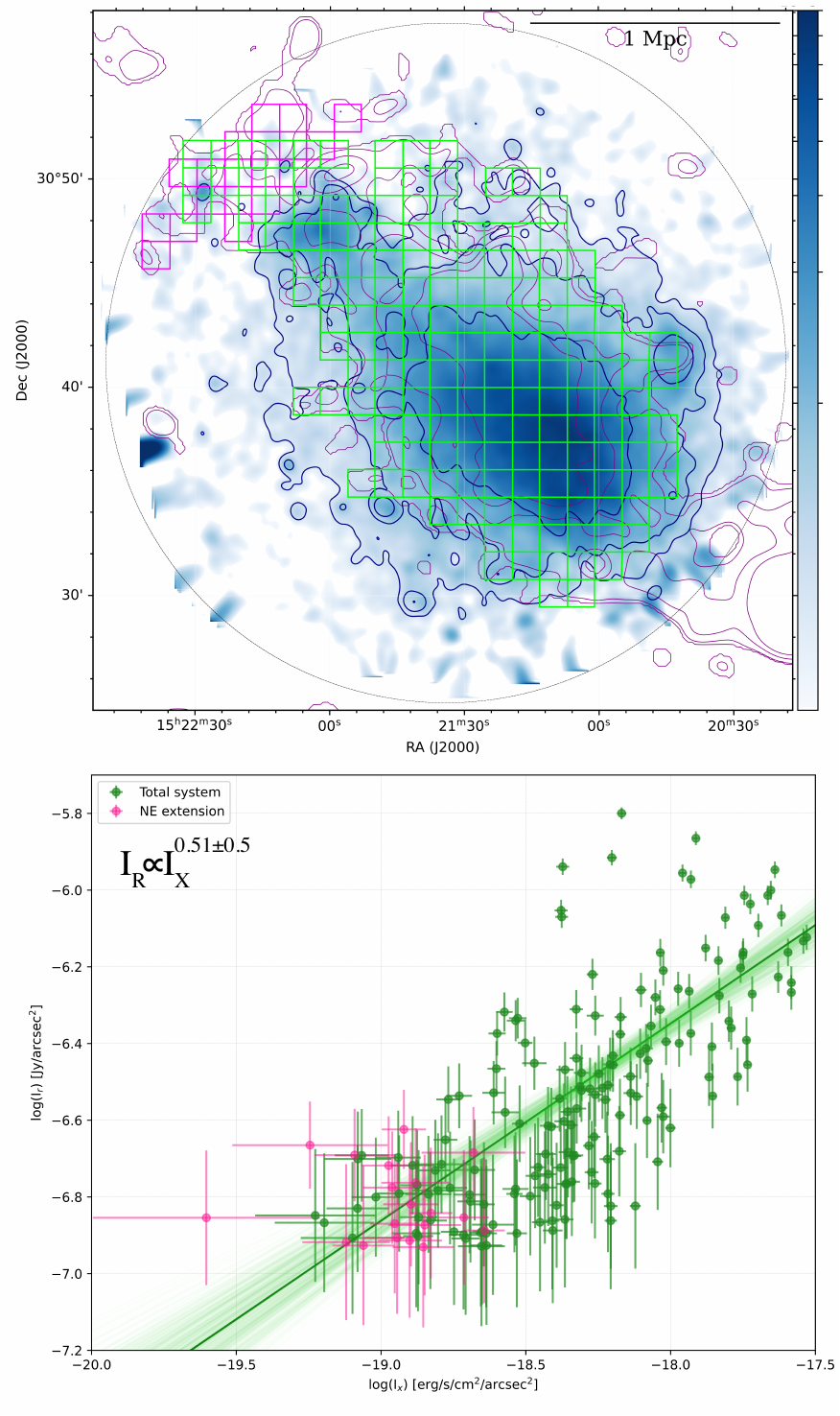}
    \caption{Point-to-point $I_{R}/I_{X}$ correlation for A2061. \textit{Top panel}: Colourscale is the \textit{XMM-Newton} surface brightness smoothed image, after subtraction of point-sources. Radio contours at 144~MHz ($80''$ resolution) are shown in purple. The regions used to extract the surface brightness for the $I_{R}/I_{X}$ correlation are $80''$ wide boxes. The green boxes cover the entire object, while magenta boxes are only over the radio emission beyond the X-ray plume. \textit{Bottom panel}: Radio/X-ray surface brightness correlation. The green solid line indicates the best-fit power-law relation for the green points. The slope of the green correlation is reported in the top right corner. The magenta points, relative to the NE extension, show no correlation. 
    }
    \label{fig:correlation}
\end{figure}
We perform a similar study of the X-ray profile of A2061 with \textit{XMM-Newton} data to check whether there is an excess emission corresponding to the radio excess.

We extracted the X-ray profile following the radio emission, starting from the same center as the radio halo and covering the X-ray plume. \textit{XMM-Newton} observations field of view extends beyond the X-ray plume, allowing a comparison between the radio emission and the X-ray in the NE extension. We extract the X-ray profile also in a different sector, slightly offset from the plume and reaching the same distance from the center. 
The extracted X-ray band radial profile is shown in \Fig\ref{radial_profile} (right panels). We note how the profile extracted in the NE extension shows an evident 'bump' at a distance of $\sim11'$ from the halo center, which corresponds to the emission from of the X-ray plume. Interestingly, beyond the X-ray plume the profile shows an excess of emission with respect to the background level at larger radii. Compared with another sector, we find that in an offset direction from the NE extension the X-ray profile is monotonically decreasing and, at the same radii, is consistent with the background level. In the scenario where the X-ray plume is residual thermal gas left behind by an infalling group \citep{marini2004}, the radial profile shows that the excess is likely not connected with this gas, since we see a clear drop of the emission brightness between the plume and the extension.
This excess might instead indicate the presence of thermal filamentary gas, however the limited field of view of the X-ray image does not allow to investigate the properties of the emission at larger distances from the center. An additional \textit{XMM-Newton} pointing covering the separation between the two cluster, will help with a definitive conclusion on the nature of this trend. The infalling group-scenario will be further discussed in \Sec\ref{sec:discussion}.

\subsection{Point-to-point analysis}\label{sec:pointopoint}

We are interested in studying the link between the diffuse radio emission and X-ray emission of the thermal plasma, that are found to be spatially aligned at the center of galaxy clusters. Past works have shown that this link can be described by a correlation of the form $I_{R}\propto I_{X}^{b}$ \citep[e.g.][]{govoni2001, rajpurohit2021b}. The slope of this correlation between the X-ray and the radio surface brightness of diffuse emission in galaxy clusters provides important information on the acceleration mechanism at work \citep[for a review, see][]{feretti2012, brunetti2014}. Moreover, the spatial distribution of the correlation can reveal the presence of different environments and emission powered by different physical mechanism, as shown in \cite{bonafede2022} for the radio halo in the Coma cluster, in \cite{rajpurohit2023} for Abell 2256 and in \cite{biava2021} for the mini-halo in RXC J1720+2638. Several studies investigated disturbed galaxy clusters, where generally the correlation results to be sub-linear \citep[see e.g.,][] {hoang2019,bonafede2022,riseley2022,riseley2024}, implying a stronger decline of the X-ray emission than the non-thermal surface brightness. In the case of A2061, we can extend the investigation of the correlation to the NE extension. 

In order to investigate the correlation, we used the $80''$ resolution, compact source-subtracted radio maps at 144~MHz and the point source-subtracted \textit{XMM-Newton} image.
We covered the entire \textit{XMM-Newton} field-of-view of A2061 with a grid of square boxes of $80''$, therefore covering the cluster and the NE extension. We only consider the boxes where the radio surface brightness is above the $2\sigma_{rms}$ level. For comparison, we also perform the analysis over the NE extension area only. The result for the $I_{R}/I_{X}$ correlation are shown in \Fig\ref{fig:correlation}. 

In the case of the total system, the two components appear mildly correlated, with Spearman ($r_{S}$) and Pearson ($r_{P}$) coefficients of $r_{S}=0.73$ and $r_{P}=0.71$, respectively. Considering the NE extension region, where we only have 19 points, we do not find evidence of correlation, with $r_{S}=r_{P}=-0.06$. 
We can quantify the slope of the correlation in the entire system fitting a power-law relation in log-log space in the form:
\begin{equation}
    \log(I_{R})=a + b\log(I_{X}).
\end{equation}
To perform the fit we use the Bayesian regression MCMC method implemented in \texttt{Linmix} \citep{kelly2007} as recently done in, e.g., \cite{riseley2024, riseley2022, rajpurohit2021b, rajpurohit2021c}. The fitting procedure yields a best-fit slope of $b=0.51\pm0.05$, showing an overall sub-linear correlation. If we remove the points over the NE extension from the total system, we find very similar correlation and best fit, with a slope of $b=0.54\pm0.05$. The connection between thermal and non-thermal emission over the NE extension area remains unclear. It is possible that a better X-ray coverage of the inter-cluster region could reveal a mild correlation. We discuss these findings in relation of three different scenarios to explain the NE extension emission.

\subsection{The radio halo and the NE extension}\label{sec:discussion}
\begin{figure*}[h!]
    \centering
    \includegraphics[width=1\linewidth]{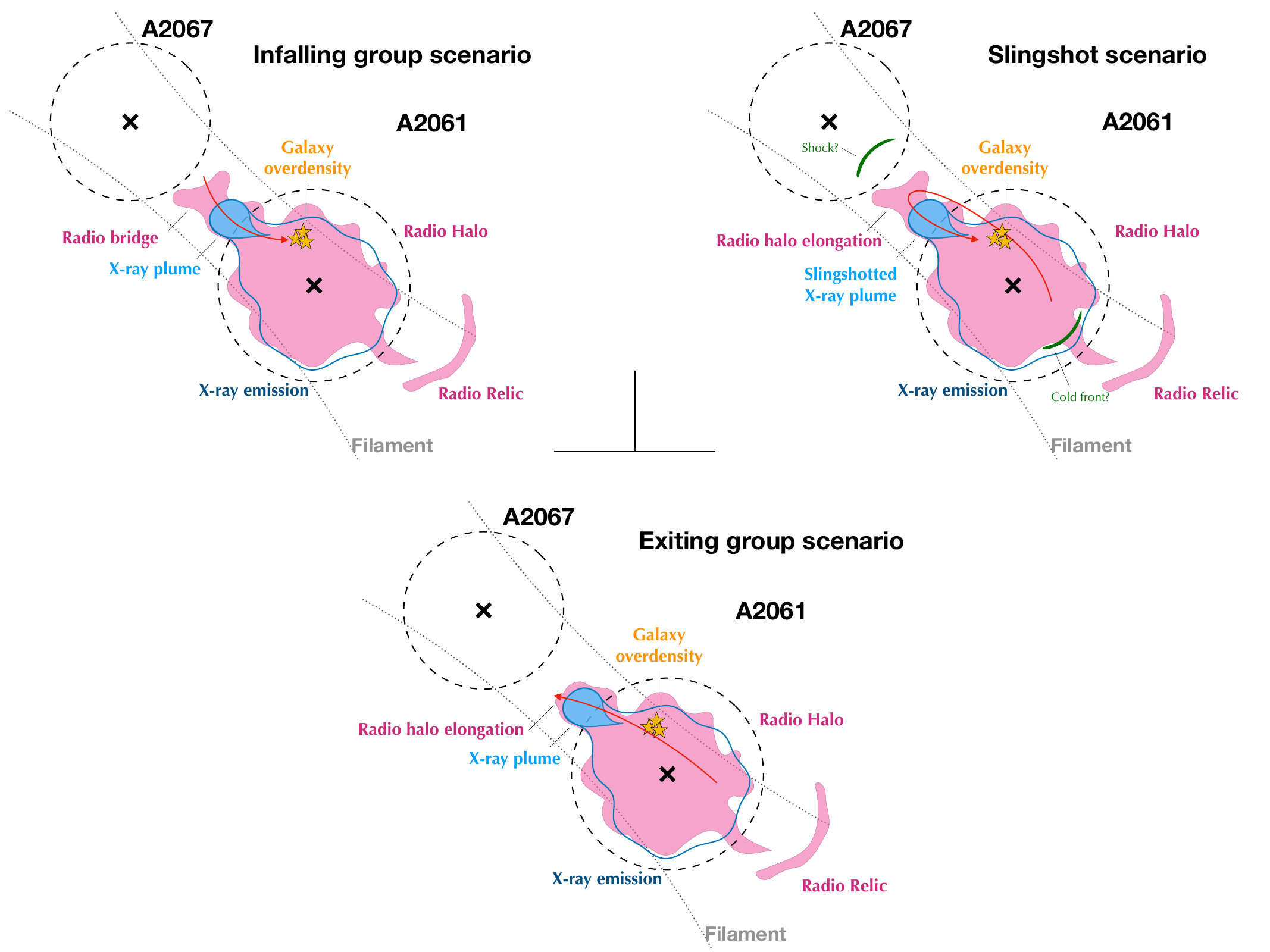}
    \caption{Schematic representation of the three different dynamical scenarios investigated in this work to explain the NE extension emission.
    }
    \label{scenarios}
\end{figure*}

The LOFAR observations presented in this work are able to detect both the giant radio halo, the radio relic and the trail in A2061, recently confirmed in \cite{botteon2022}, with the additional detection of large-scale diffuse emission extending towards A2067. 
In \cite{farnsworth2013}, they observe diffuse emission in A2067 offset from the X-ray peak, tentatively classified as a radio relic, that is connected to the radio halo in A2061 with a filamentary feature. The elongation is present also in \cite{botteon2022} images. Our new LOFAR observations confirm the presence of such emission, extending for $\sim800$~kpc between the radio halo in A2061 and A2067, of which an additional $\sim350$~kpc component in the NE direction is unveiled with respect to the previous LoTSS observations \citep{botteon2022}. However, this emission does not fully connect the two galaxy clusters. The low resolution ($570''\times560''$) and limited sensitivity of the GBT-NVSS observations most likely resulted in a blending of the tails of unresolved radio sources present in A2067 - that were not included in the NVSS subtraction process - which was interpreted as a radio relic. We are able to resolve the sources, and the residual tails of the AGN are also shown in our maps, but there is no evidence of other extended emission in this cluster. Even re-convolving our images to the GBT resolution, the filamentary emission is still not connecting the two galaxy clusters. There are few cases where features of diffuse emission of uncertain nature are detected between pre-merging galaxy clusters that do not show the presence of radio halos \citep{gu2019, kurahara2023}. In these cases, the emission is either resulting from present radio sources revived by the merger activity \citep{gu2019}, or a radio relic of peculiar morphology \citep{kurahara2023, omiya2023}. In both these cases, the lack of radio halos in both clusters help the classification, as there is no mixed diffuse emission. In the NE extension, the emission is connected to the radio halo in A2061 and with our analysis we are confident to exclude that the emission is related to a phoenix or the blending of AGNs in the field.

\begin{figure}[h!]
    \centering
    \includegraphics[width=1\linewidth]{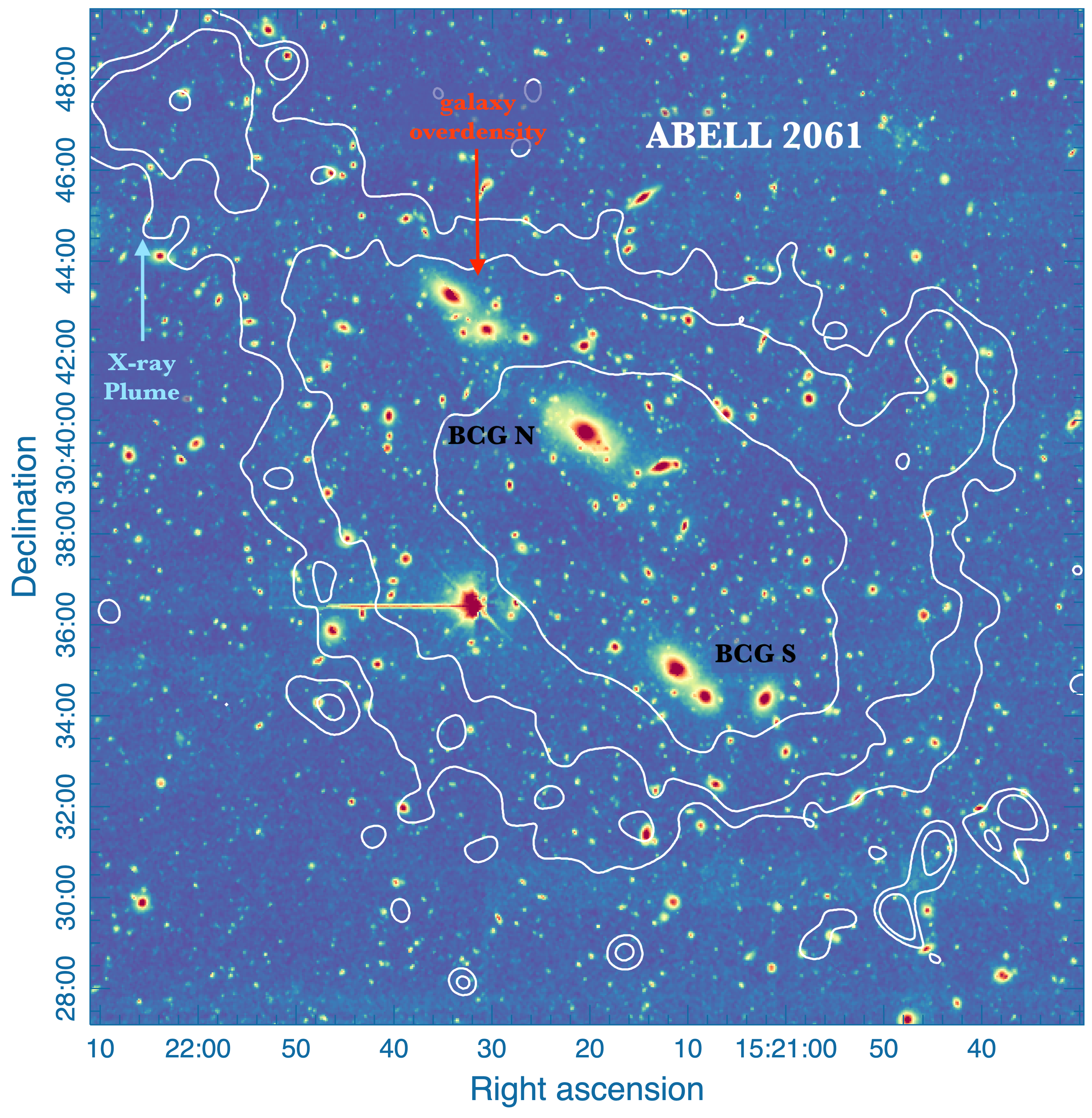}
    \caption{Optical image of A2061. Colourscale is Dark Energy Spectroscopic Instrument (DESI) Legacy Survey optical image of the galaxies in A2061. \textit{XMM-Newton} contours are overlaid in white. In black we mark the location of the two BCGs, in red the location of a galaxy overdensity attributed to the infall of a group of galaxies \citep{marini2004}, and in cyan the location of the X-ray plume.
    }
    \label{fig:desi}
\end{figure}

Here, we discuss three possible scenarios to explain the NE extension emission (see \Fig\ref{scenarios}):
\begin{itemize}
    \item \textbf{Infalling group scenario:} initially, the extension towards the NE was interpreted as possibly related to the X-ray plume. According to \cite{marini2004}, the plume can be attributed to a group of galaxies arriving from the NE, infalling in A2061. As a consequence of the impact, the galaxies of the group precede its intracluster medium, and are observed as an overdensity in the bidimensional galaxy distribution \citep[see \Fig\ref{fig:desi}, also][]{marini2004}. In this case, the NE extension could be classified as a radio bridge.

    \item \textbf{Slingshot scenario:} a slingshot X-ray tail can be formed as a (secondary) sub-halo moves away from the (primary) cluster center toward the apocenter of its orbit \citep{poole2006, markevitch2007, sheardown2019}. In this scenario, a galaxy group might have already completed a first passage, and it is now approaching for a second infall from NE. Therefore, in this case the X-ray plume can be explained as the slingshotted gas tail of the re-entering group, and the NE extension as an elongation of the radio halo caused by the passage of the group. There are only few examples of observational features in galaxy clusters that can be explained by the slingshot effect, for example in Abell 168 \citep{hallman2004}, in the Fornax Cluster \citep{sheardown2018}, and in the Coma Cluster \citep{lyskova2019, lal2022}. In this scenario, simulations presented in \cite{sheardown2019} predict the possible presence of a shock at large distances behind the secondary tail, and/or the presence of a cold front in the primary cluster, attributed to the sloshing in the core caused by the passage of the secondary (Akamatsu et al., in prep.).
    
    \item \textbf{Exiting group scenario:} in this case, a sub-halo might be moving from the South-West direction, crossing the cluster and causing the elongated shape in the radio halo as it exits on the opposite side \citep[see e.g.][]{donnert2013, beduzzi2024}. From simulations results, in this scenario the extended radio emission from the radio halo would be, at best, completely co-spatial with the X-ray plume.
    
\end{itemize}

Our observations, compared with X-ray maps, show that the radio emission is far more extended towards A2067, beyond the X-ray plume, as seen both by ROSAT (\Fig\ref{fig:composite}) and \textit{XMM-Newton} (\Fig\ref{radial_profile}, right panels), disfavouring the exiting group scenario. The radio surface brightness profiles presented in \Sec\ref{sec:radioradial} show that the surface brightness measured outside $~2.5$ e-folding radii in the NE direction significantly deviates from the analytical radio halo profile (\Fig\ref{radial_profile}, left panels). While some deviations between model and real data in disturbed radio halos are expected \citep{botteon2023}, the excess appears to be systematic at large radii.
Moreover, the X-ray radial profile in the same direction (\Sec\ref{sec:xrayradial}) shows a similar trend beyond the X-ray plume location, where a significant excess over the background is detected (\Fig\ref{radial_profile}, right panels). Theoretically, in the early stages of a merger, we can expect enhancements of the gas density between approaching clusters, which, in turn, could be detected as an X-ray excess \citep{machado2024}. Since this source extends along the merger axis, it may be related to the underlying filament connecting the two galaxy clusters. Numerical simulations indicate that even during the early phases of cluster interactions, weak shocks (Mach $\sim 2-3$, e.g., \citealt{ha2018, vazza2019}) and transonic turbulence \citep{iapichino2011, beduzzi2024} can be triggered in the inter-cluster medium following the compression and accretion of smaller sub-clusters. Therefore, the filament volume can undergo a significant level of Fermi II re-acceleration by solenoidal turbulence motions injected in the early stages of the merger process, provided that the medium hosts a pool of mildly relativistic electrons ($\gamma\sim10^{3}$), which is supported by recent simulations \citep{beduzzi2023}. We have observational evidence of radio bridges between galaxy clusters in the early stages of merger \citep{botteon2018, botteon2020, govoni2019, dejong2022,pignataro2024a, pignataro2024b}, where the dynamics of the collapse drives transonic turbulence in the compressed intercluster gas \citep{brunetti2020, nishiwaki2024}. The weak shocks in the bridge can also compress turbulent, re-accelerated electrons and magnetic fields, leading to increased radio brightness in that region. The same acceleration processes could be at play in this case, with the additional amplification of turbulence brought by the merger of groups \citep{brunetti2020}, as we would expect in the infalling group scenario.

The case in which one of the two clusters does not host a radio halo, opens up a question on the definition of radio bridge. In the two most spectacular cases of radio bridges \citep{govoni2019, botteon2020}, the bridge fully connects the two radio halos at the center of the galaxy clusters. In this scenario, the lack of a radio halo in A2067 is likely the reason we observe a `gap' of emission between the two clusters. Nonetheless, these system properties, such as the projected separation and the average  emissivity of the diffuse emission, closely resembles the other two discovered bridges, as summarized in \T\ref{tab:comparisonbridge}. In this view, the emerging population of low-surface brightness bridges might represent a more ubiquitous phenomena in merging clusters of galaxies, compared to the appearance of classical radio halos in their clusters. In fact, radio halos additionally require the presence of mergers within clusters \citep{brunetti2007, brunetti2009, vanweeren2019}, on top of the large-scale accretion motions that instead may power the radio emission in bridges \citep{brunetti2020}.

However, the radio/X-ray correlation investigated in \Sec\ref{sec:pointopoint} does not provide conclusive evidence. The similar scaling found including and excluding the NE extension emission seems to indicate a trend where the emission traces an extension of the radio halo, which could be the case in the slingshot scenario. In contrast, in the scenario of filamentary gas, we would expect a level of X-ray/radio correlation, as it has been found for the radio bridges in A399-A401 \citep{dejong2022} and A1758N-S \citep{botteon2020}. In both cases, the X-ray observations allowed a full coverage of the bridge, and they are able to create a more extended grid in the bridge region. They both showed good correlation over the bridge, and nearly flat slopes for the correlation \citep[$b\sim0.25,$][]{dejong2022}. 
In this case the limited field of view of the \textit{XMM-Newton} observation does not allow to reach a firm conclusion on the presence and nature of thermal gas in the inter-cluster region. Additionally, assuming a slingshot scenario, for a system that has likely underwent a history of merger with more than one sub-halo, as it is suggested by the disturbed morphology and the presence of multiple BCGs in A2061 (\Fig\ref{fig:desi}, \citealt{hill1998}), it is not straightforward to attribute all dynamical features in the primary core to only a secondary halo \citep{sheardown2019}.

As a result of our current analysis, we are not able to firmly favour the infall group scenario or the slingshot scenario. Therefore, the classification of the NE extension remains open, as it could be described either as a radio bridge or a natural extension of the radio halo. For the NE extension, outside A2061 $R_{500}$ we measure an integrated flux density of $S_{144}^{2\sigma}=52\pm5.0$ mJy. To confirm the classification, it would require a spectral analysis with multifrequency radio detections, and a complementary X-ray/radio point-to-point analysis with a X-ray pointing covering the entire inter-cluster separation, as well as simulations that can help distinguish between the two most likely dynamical scenarios.

\begin{table}[]
\renewcommand{\arraystretch}{1.3}
\centering
\resizebox{\columnwidth}{!}{%
\begin{tabular}{ccccc}
\textbf{System}      & \textbf{$\mathbf{\langle z \rangle}$} & \textbf{\begin{tabular}[c]{@{}c@{}}$\mathbf{\langle \epsilon_{144} \rangle}$\\ {[}erg s$^{-1}$ Hz$^{-1}$cm$^{-3}${]}\end{tabular}} & \textbf{\begin{tabular}[c]{@{}c@{}}D$_{centers}$\\ {[}Mpc{]}\end{tabular}} & \textbf{\begin{tabular}[c]{@{}c@{}}$\mathbf{M_{tot}}$\\ {[}M$_{\odot}${]}\end{tabular}} \\ \hline \hline
\textbf{A2061-A2067} & 0.076                                 & 4.5$\times10^{-43}$                                                                                                                & 2.5                                                                        & 5 $\times10^{14}$                                                                       \\
\textbf{A399-A401}   & 0.07                                  & 8.6$\times10^{-43}$                                                                                                                & 3                                                                          & 1.5 $\times10^{15}$                                                                     \\
\textbf{A1758N-S}    & 0.28                                  & 4.1$\times10^{-43}$                                                                                                                & 2                                                                          & 3 $\times10^{15}$                                                                       \\ \hline
\end{tabular}%
}
\smallskip
\caption{Comparison between the radio bridges discovered in A399-A401 \citep{govoni2019} and A1758N-S \citep{botteon2020} and the properties of the NE extension emission between A2061 and A2067. The first column is the average redshift of the system, the second column is the average radio emissivity at 144~MHz, computed assuming a cylindrical volume between the two clusters, the third column is the projected distance between the centers of the clusters, and the last column is the total mass of the system. }
\label{tab:comparisonbridge}
\end{table}
\subsection{The radio relic and trail}\label{sec:relic}

Finally, we discuss the other sources of diffuse emission in A2061, the radio relic and the trail.

The presence of the radio relic was confirmed by WSRT observations at 1.38 and 1.7~GHz \citep{vanweeren2011}. In their work, they fit the spectrum of the radio relic with a single power-law between 327~MHz, 1.38 and 1.7~GHz. However, the 327~MHz flux measurement is uncertain \citep{rudnick2009}.
To better constrain the spectral index, we measure the flux density of the radio relic at 144~MHz from our observations. We used an image at resolution of $30''$, similar to the resolution of the high frequency images ($32''\times15''$). At this resolution, we measure for the relic a largest-linear-size (LLS) of $\sim790$ kpc inside the $5\sigma_{rms}$ level. The integrated flux density of the relic was computed in the same polygonal region encompassing the $3\sigma$ contour of the diffuse emission at 1.38 GHz used in \cite{vanweeren2011}. We measure a flux density of $S_{144}=220\pm22$ mJy, and find a spectral index $\alpha=-0.92\pm0.05$, consistent with the one derived by \cite{vanweeren2011}.

Another interesting source of diffuse emission in this system is the `trail' of emission connecting the radio halo to the radio relic. This feature is extended with a LLS$\sim700$ kpc. Similar features of connected diffuse emission were also detected in several systems, most recently in A1550 \citep{pasini2022}, in A3667 \citep{carretti2013,degasperin2022}, and others \citep[see e.g.][]{kim1989, vanweeren2012,rajpurohit2018, bonafede2018, degasperin2020b}. The most striking example of diffuse emission connecting a radio halo and a radio relic is the Coma cluster bridge \citep{bonafede2021}, with a LLS of 940~kpc. These trails of diffuse emission are probably related to post-merger generated turbulence that re-accelerates particles previously accelerated by shocks \citep{vanweerenbrunetti2016,bonafede2021, pasini2022}. It is likely that the trail of emission in this system could have the same origin. For the trail between the radio halo and the relic in A2061 we measure an integrated flux density of $S_{144}^{2\sigma}=49\pm5$ mJy inside the 2$\sigma_{rms}$ contour level. 
All the radio quantities measured in this work for the sources in A2061 and the bridge are summarized in \T\ref{tab:flux_summary}.

\begin{table}[]
\renewcommand{\arraystretch}{1.3}
\centering
\resizebox{\columnwidth}{!}{%
\begin{tabular}{ccccc}
                      & \textbf{\begin{tabular}[c]{@{}c@{}}LLS \\ {[}kpc{]}\end{tabular}} & \textbf{\begin{tabular}[c]{@{}c@{}}Flux \\ {[}mJy{]}\end{tabular}} & \textbf{\begin{tabular}[c]{@{}c@{}}P$_{144}$ \\ {[}W/Hz{]}\end{tabular}} & \textbf{\begin{tabular}[c]{@{}c@{}}Spectral index$^{*}$\\ {[}$\alpha${]}\end{tabular}} \\ \hline \hline
\textbf{NE extension} & 800                                                               & 52$\pm$5.0                                    & (8.1$\pm$0.8)e23                                                         & -1.4                                                                             \\
\textbf{Radio Halo}   & 1300                                                              & 265$\pm$26                                                         & (4.1$\pm$0.4)e24                                                         & -1.3                                                                             \\
\textbf{Radio Relic}  & 790                                                               & 220$\pm$22                                                         & (3.3$\pm$0.3)e24                                                         & -0.92$\pm$0.05                                                                              \\
\textbf{Trail}        & 760                                                               & 49$\pm$5.0                                                         & (7.5$\pm$0.8)e23                                                         & -1.4                                                                             \\ \hline
\end{tabular}%
}
\smallskip
\caption{Summary of the radio quantities measured for the diffuse sources in A2061. The NE extension and trail quantities are measured on the $2\sigma_{rms}$ contours of the $80''$ resolution map. The radio halo quantities are measured on the $6\sigma_{rms}$ contours of the $80''$ resolution map. The radio relic quantities are measured on the $30''$ resolution map (see text).  $^{*}$Value of the spectral index assumed for the radio power calculation (except the radio relic, which is derived in \Sec\ref{sec:relic}). The assumed value for the bridge and trail is based on literature works \citep{pignataro2024a, bonafede2021}. }
\label{tab:flux_summary}
\end{table}

\section{Conclusions}

Using deep LOFAR HBA observations, we detected diffuse extended emission at 144~MHz between the dynamically interacting galaxy clusters A2061 and A2067, inside the Corona Borealis Supercluster.
A2061 is a highly dynamically disturbed system, showing several sources of diffuse emission. The radio halo and the radio relic were already classified with previous radio observations \citep{farnsworth2013, vanweeren2011, botteon2022}. With new 16 hours of observations at 144~MHz, we are able to detect a source of emission extending for $\sim 800$ kpc from the radio halo in A2061 towards A2067. We find no evidence of a radio halo in A2067. The results of our radio and X-ray analysis on the NE extension can be summarized as follows:

\begin{itemize}
    \item From the study of the radio surface brightness radial profile, we find that the NE extension emission deviates from the analytical model for the radio halo in A2061. 
    \item From the study of the X-ray surface brightness radial profile, we find an excess of emission over the background level, which is co-spatial with the radio emission in the extension. This excess is not measured in a different direction, at the same radii, where we measure the expected decline of the X-ray profile.
    \item From the point-to-point radio-X correlation, we find mild correlation between thermal and non-thermal emission, measured over the total system. The NE extension only does not show any clear correlation.  
\end{itemize}

From these results, we investigate three different dynamical scenarios that can help classify the emission in the NE extension. In the exiting group and slingshot scenario, the diffuse emission can be classified as an elongation of the radio halo, while in the infalling group scenario, the emission can be classified as a radio bridge. Our analysis disfavour the exiting group scenario, while can not firmly confirm the infalling group or slingshot scenarios.
The system closely resembles the two other systems were the presence of a radio bridge is confirmed \citep{govoni2019, botteon2020}, yet the absence of a radio halo in one of the two clusters makes this cluster pair unique at the moment.
Further analysis, in particular the forthcoming works by Akamatsu et al. (in prep.), which will provide a robust X-ray analysis based on both Suzaku and new XMM-Newton observations, will help clarify the on-going dynamical processes in A2061, and therefore understand the nature of the diffuse emission between the two galaxy clusters.

\begin{acknowledgements}
AB acknowledges financial support from the ERC Starting Grant `DRANOEL', number 714245. FV acknowledges the support by Fondazione Cariplo and Fondazione CDP, through grant n° Rif: 2022-2088 CUP J33C22004310003 for "BREAKTHRU" project.
RJvW acknowledges support from the ERC Starting Grant ClusterWeb 804208. AB acknowledges financial support from the European Union - Next Generation EU.
\end{acknowledgements}

%
%
\bibliographystyle{aa}
\bibliography{bibliography}

\begin{appendix} 

\section{Halo-FDCA}\label{appendix}
\begin{figure}[h]
    \centering
    \includegraphics[width=1\linewidth]{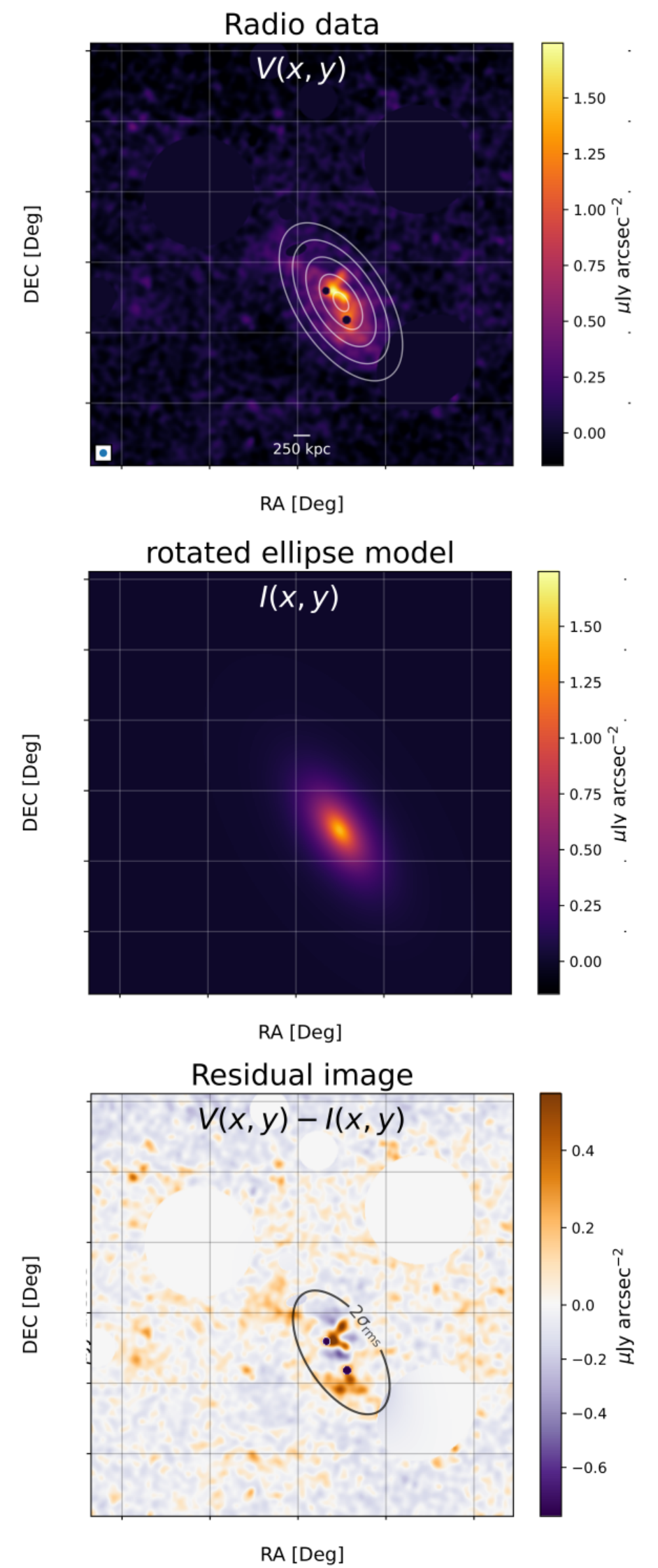}
    \caption{Result from the fitting procedure performed by \texttt{Halo-FDCA} on the radio halo in A2061. The residual from the extended radio emission in the NE extension is visible in the bottom panel.
    }
    \label{fig:appendix}
\end{figure}
\end{appendix}

\end{document}